\def\apj{{ApJ}}

\def\mnras{{MNRAS}}
\def\nat{{Nature}}

\def\simless{\mathbin{\lower 3pt\hbox
   {$\rlap{\raise 5pt\hbox{$\char'074$}}\mathchar"7218$}}} 
\def\simgreat{\mathbin{\lower 3pt\hbox
   {$\rlap{\raise 5pt\hbox{$\char'076$}}\mathchar"7218$}}} 

\documentclass[cup5b]{caps}
\usepackage{graphicx}
\usepackage{amssymb}
\usepackage{ociwsymp1e}
\HeadText{M. Colpi, M. Mapelli, and A. Possenti}

\begin{document}

\pagenumbering{arabic}

\author[]{M. COLPI$^1$, M. MAPELLI$^1$, and A. POSSENTI$^{2,3}$ 
\\$^1$Department of Physics, Universit\`a Degli Studi
di Milano Bicocca, Milan, Italy 
\\$^2$Cagliari Astronomical Observatory, Italy
\\$^3$Bologna Astronomical Observatory, Italy}

\chapter{Is NGC 6752 Hosting a Single or a Binary Black Hole?}

\begin{abstract}
The five millisecond pulsars that inhabit NGC 6752 display locations or
accelerations remarkably different with respect  to all other pulsars
known in globular clusters.
This may reflect  the occurrence of an uncommon dynamics  in the cluster
core 
that could be attributed 
to the presence of a massive perturber.
We here investigate whether a single intermediate-mass black hole, 
lying on the extrapolation of the mass ${\cal M}_{\rm BH}$ 
versus $\sigma$ 
relation observed in  galaxy spheroids, or, a less massive black hole binary
could play the requested role.

\end{abstract}

\section{The Peculiarities of NGC 6752}

NGC 6752 is a relatively close (distance $d=4.3$ kpc, Ferraro et
al. 1999) globular cluster that harbours 5 millisecond pulsars
(D'Amico et al. 2002) all displaying unexpected characteristics. 
PSR-A (a binary pulsar) holds the record of being the farthest
millisecond pulsar ever observed from the center $C_{\rm grav}$ of a
globular cluster (at a distance of $\approx$ 3.3 half mass radii).
PSR-C, an isolated pulsar, ranks second in the list of the most
offset pulsars (being at a distance of 1.4 half mass radii from
$C_{\rm grav}$). PSR-B and PSR-E (located within the
cluster core) have very high {\it negative} spin derivatives.

If the {\it negative} derivatives are ascribed to the overall effect of
the cluster potential well (as customarily assumed), the resulting
central mass to light ratio $(M/L)$ is much larger (Ferraro et al. 2003)
than that published $\sim 1.1$ (Pryor \& Meylan 1993) and at least {\it
twice} as large as the typical value $M/L\sim 2-3$ observed in the
sample of the core collapsed clusters (Pryor \& Meylan 1993). This
peculiarity could be explaind with $\sim 1000~{\rm M_\odot}$ of low
luminosity unseen matter enclosed within the central 0.08 pc of the
cluster (Ferraro et al. 2003). In this scenario even the very high {\it
positive} $\dot{P}$ (the third ever observed in a globular cluster
pulsar) of PSR-D could be easily explained (even though in this case it
could also be intrinsic).  Nevertheless, one could argue that the negative
values of $\dot{P}$ of PSR-B and PSR-E are influenced by the
gravitational pull of some local perturber, such as nearby passing stars
or even more massive objects (Ferraro et al. 2003).

Interestingly, {\it all} the millisecond pulsars
of NGC 6752 manifest peculiarities. In addition 
this cluster presents a complex stellar density profile 
(Ferraro et al. 2003). The
combination of these facts strongly suggests that some uncommon
dynamics affects the core and the halo of NGC 6752.
 
\section{Black Hole(s) in the Core of NGC 6752?}

Recently, it has been conjectured (Colpi, Possenti \& Gualandris 2002)
that NGC 6752 may hide in its core a rather exotic binary comprising
two black holes (BHs) with masses $\approx 10-100 M_{\odot}.$ Born
from the most massive stars (Heger et al. 2003) these BHs may grow 
by cannibalizing other BHs through mergers 
(Miller \& Hamilton 2002) 
or occasionally capturing stars. Binary BHs may have escaped the cluster as
a result of repeated exchange interactions with BHs, when the cluster
was younger (Sigurdsson \& Hernquist 1993; 
Portegies Zwart \& McMillan 2000).  But, at least one binary, the last
and  heaviest
($\ge 50 M_{\odot}$), could have been  retained in the cluster
core (Miller \& Hamilton 2002). NGC 6752 seems a primary candidate for 
the search of such exotic binary:  Colpi, Possenti
\& Gualandris (CPG hereafter) suggested to use PSR-A, the 
binary millisecond pulsar in the halo of NGC 6752, 
to infer its presence, starting from considerations on the uncommon
location of PSR-A.
 
CPG first explored the possibility that PSR-A descends from a
primordial binary, either born in the halo, or in the core (in the
latter case it would have been ejected as a consequence of a natal
kick).  Their analysis however led to discard both these hypotheses as well
as that of a scattering or exchange event off a core star,
given the constraints imposed by the binary nature
of PSR-A.  CPG thus
conjectured that {\it a binary of two BHs} can be massive enough to provide
the mechanism for propelling PSR-A into its current halo orbit,
at an acceptable event rate.  

Prompted by the  evidence of 
an  unusually high $M/L$ ratio in the core of NGC 6752 (Ferraro et al. 2003), 
we wondered whether the position of PSR-A could be explained  in 
an alternative way. We thus started to consider 
the possibility of ejecting  PSR-A into the cluster periphery 
following a dynamical encounter with 
{\it a central intermediate-mass 
BH}.  According to an old suggestion by Frank \& Rees
(1976) stars in the {\it cusp} of a central BH can eject other
stars approaching nearby (Lin \& Tremaine 1980): PSR-A may have just 
experienced  such a collision. 

On a theoretical ground, 
formation of very
massive stars via stellar collisions (Portegies Zwart \& McMillan
2002) may ultimately lead to the presence of single intermediate-mass BHs
($\approx 1000 {\rm M_\odot}$) in the cluster cores
(see also Mouri \& Taniguchi 2002).
Recently, HST/STIS
observations of the globular cluster G1 in M31 (Gebhardt et al. 2002)
and of M15 in the Milky Way (Gerssen et al. 2002) have provided clues
for the presence of a central BH.  Surprisingly and
even more puzzlingly, the two postulated intermediate-mass 
BHs in G1 and M15 seem to lie
just along the extrapolation of the BH  mass versus dispersion
velocity ($\cal M_{\rm BH}-\sigma$) relation obeyed by the supermassive BHs
in galaxy spheriods (Ferrarese \& Merritt 2000; Gebhardt 2000).  We
thus assume that the putative central BH in NGC 6752 lies on the
$\cal{M}_{\rm BH}-\sigma$ relation: taking the recent spectroscopic
value of $\sim 7$ km s$^{-1},$ (Xie et al. 2003) for the line-of-sight
dispersion velocity $\sigma$ of NGC 6752, we deduce a mass ${\cal{M}_{\rm
BH}}\sim 500 M_\odot.$

We report on an ongoing study of simulated binary-binary encounters in
the aim at {\bf [i]} investigating the aforesaid hypothesis of
scattering of the binary pulsar PSR-A off the {\sc cusp of a single
black hole }, and at {\bf [ii]} constraining, in the case of 
PSR-A scattering off a {\sc black hole binary},
the possible masses of the latter. 

\section{Four-Body Encounters}

PSR-A orbits around a companion (probably degenerate) star (hereafter
COM) of $m_{\rm COM}=0.2~{\rm M_\odot}$ at a distance of 0.0223 AU
($P_{\rm{orb,PSR-A}} = 0.86$ days), with an eccentricity $e_{\rm
PSR-A}\leq{} 10^{-5}.$ PSR-A has experienced a phase of recycling (and
of orbital circularization) that has driven the neutron star to spin
at the observed period of 3.27 ms.

\subsection{Ejection by a [BH + Cusp Star] binary}
\label{BHsingle}

In our scenario, the single hypothetical BH in NGC 6752, of
${\cal{M}}_{\rm BH}\sim 500~{\rm M_\odot},$ is
surrounded by a swarm of bound stars belonging to the {\it cusp}, a
region extending up to a distance $r_{\rm BH}\approx G{\cal M}_{\rm
BH}/3\sigma^2\sim 0.02 {\cal M}_{\rm BH,500}/\sigma^2_{7}$ pc,
where $\sigma_{7}$ is the central velocity dispersion in units of 7 km
s$^{-1}.$ We select a cusp star (hereafter CS) of $1 {\rm {\rm M_\odot}}$
tightly bound to the BH moving on a Keplerian orbit with mean
separation $a_*$ of one AU, and eccentricity 0.7 ($P^*_{\rm orb}=16$ days).  
CS is well inside the critical
radius $r_{\rm crit}$ at which the stellar flow (percolating across
the loss cone) peaks and where significant changes of the integrals of
motion occur over one orbital period due to relaxation (Shapiro \&
Lightman 1977). This guarantees the stability of its orbit also  
during the characteristic time of the encounter $\tau_{\rm enc}\sim
{\sqrt{a_*b}}/\sigma$ where $b$ is the impact parameter of the encounter.
In general, the following
inequalities hold: $P_{\rm{orb,PSR-A}}< P^*_{\rm
orb}\simless \tau_{\rm enc}<\tau_{\rm rel}$ where $\tau_{\rm rel}$ is the
relaxation time inside the cusp.

\begin{figure}
\includegraphics[width=5.5cm]{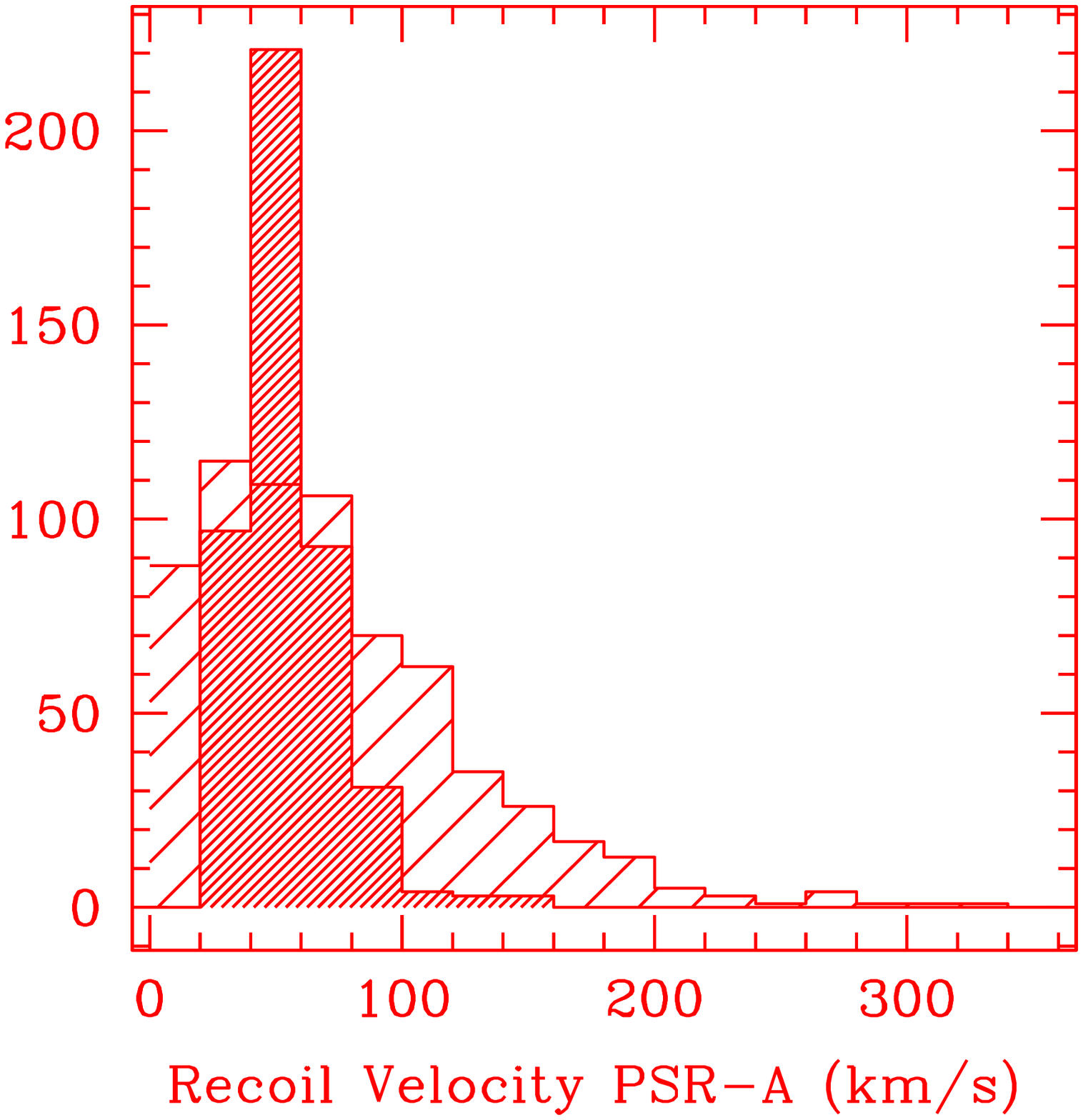}
\includegraphics[width=5.5cm]{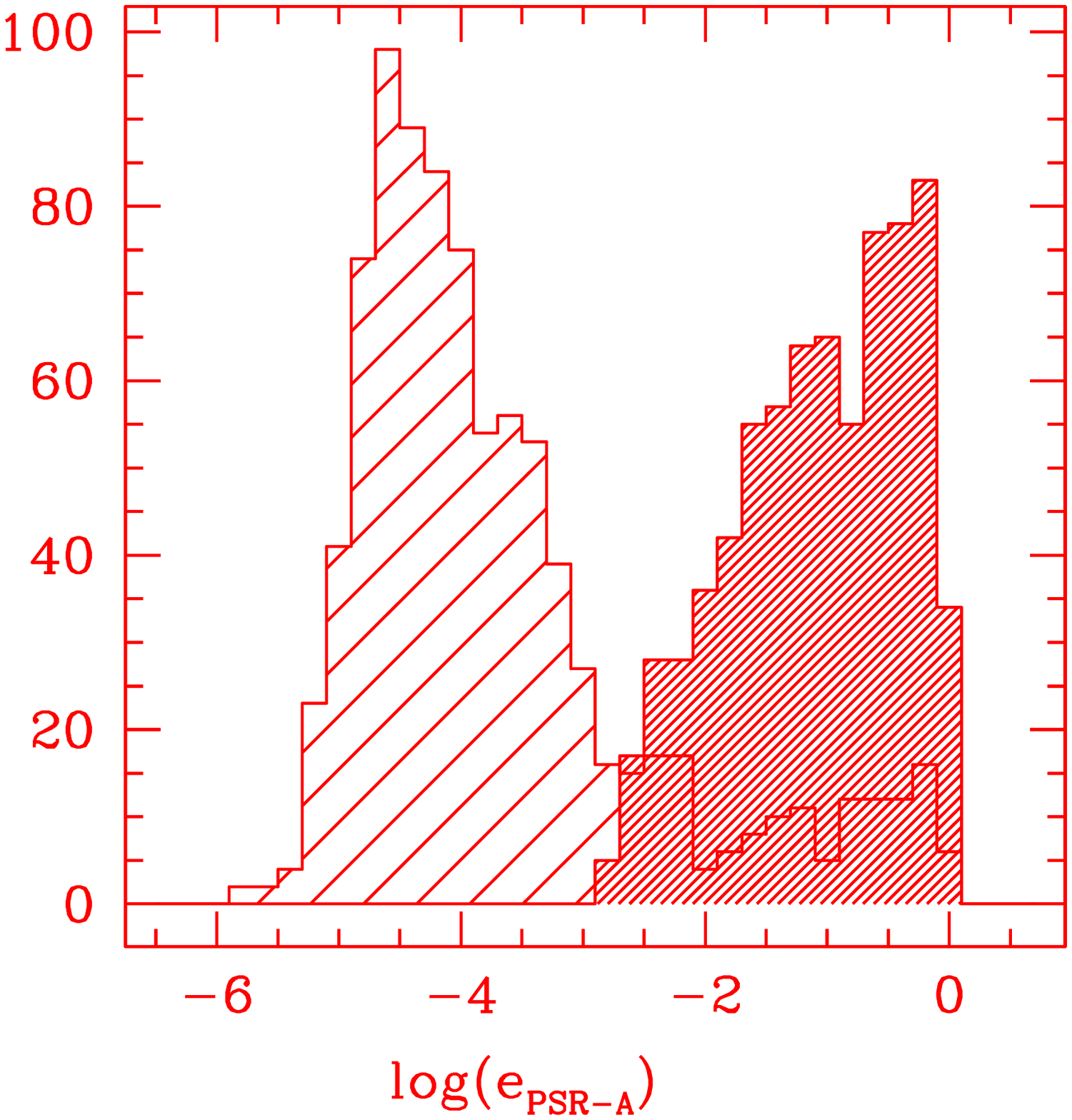}
\caption{Distribution of the ejection velocities ({\it left
panel}) and post-encounter 
eccentricities $e_{\rm PSR-A}$ ({\it right panel}) for the
binary pulsar PSR-A scattering off the [BH,CS] binary ({\it filled
histogram}) and off the [BH,BH] binary with masses (10${\rm
{\rm M_\odot}}$,50${\rm {\rm M_\odot}}$) ({\it hatched histogram}).}
\label{sample-figure}
\end{figure}

The scattering of the binary pulsar [PSR-A,COM] impinging off
the binary [BH,CS] starts from nearly parabolic orbits.
Figure 
\ref{sample-figure} ({\it filled histogram}) shows the velocity and
eccentricity distribution of the end-states of [PSR-A,COM] from 
1000 four-body encounters simulated according to the recipes of Hut
\& Bahcall (1983).  The ejection velocity necessary to propel
[PSR-A,COM] into its halo orbit (see CPG for details) is $\sim 30$ km
s$^{-1},$ and this is  in accordance with the mean velocity
$\langle V\rangle_{\rm PSR-A}$ extracted from the simulations (see
in addition Table \ref{sample-table}, last column; note that increasing $a_*$
introduces a drift toward lower mean velocities). 
We find however that the binary pulsar emerges after the 
scattering with a high eccentricity, typically
around 0.01 (right 
panel of Fig. \ref{sample-figure}).  If the circularization time exceeds 
the dynamical friction time of PSR-A in the halo ($\sim 10^9$ Gyr), the
ejection  due
to a relatively massive single BH seems incompatible with the observed
eccentricity $e_{\rm PSR-A}\leq{} 10^{-5}$ of PSR-A. 
The hypothesis of a single BH remains  still 
a possibility, if the binary hosting the pulsar was propelled into the
halo before recycling,  
when the companion star was heavier ($\sim 1~M_{\odot}$) and the binary
wider, a case that we are now investigating in more detail.
For it, the duration of the recycling phase matches with the persistence
of [PSR-A,COM] in the cluster halo according to dynamical friction,
and evolutionary constraints (Colpi, Mapelli \& Possenti 2003, in prep.).

\subsection{Ejection by a [BH + BH] binary}
\label{BHbinary}

Our simulated binary comprises two BHs of mass
$(m_{\rm BH},M_{\rm BH})$ of $(10M_{\odot},\break 50M_{\odot})$ 
having a
separation $a_{\rm BH}$ (of 1 AU) consistent with the request of survival
against escape (by recoil with other wandering BHs) and coalescence by
emission of gravitational waves (see CPG). Figure \ref{sample-figure}
({\it hatched histogram}) shows that the post-encounter velocity
distribution of the binary [PSR-A,COM] peaks around 50 km s$^{-1},$
but it appears significantly wider in the
{\it double BH} scenario than in the {\it single BH
+ cusp star} hypothesis. Both velocity distributions (left panel of
Fig. \ref{sample-figure}) have common  positions of their
peaks, and, despite the presence of a high velocity tail in the binary
BH case, $\sim 30$\% of the events fall in an interval between
20-50 km s$^{-1}$, suitable for PSR-A ejection.

Conversely, we find a remarkable difference in the post-encounter
values of the eccentricity 
(right panel of Fig. \ref{sample-figure}). In particular, in the
case of interaction of [PSR-A,COM] with a [BH,BH] binary, the 
end-state eccentricities peak around  $\sim 2\cdot 10^{-5}$ when
limiting the statistical investigation to the interval of outgoing
velocities appropriate for PSR-A ejection.  Thus, this type of
encounters seem to preserve the original circularity of the
binary pulsar.  
The difference in the post-encounter values of the eccentricity between the
BH binary case and the single one
may be ascribed to 
gravitational focusing. In fact, in the interaction of [PSR-A,COM] on
a binary BH, the resultant gravitational stresses onto the
incoming binary are reduced (due to the smaller total mass of the
system) and somehow averaged (due to the simultaneous presence of two
BHs), thus causing less damage to the internal orbital parameters of
PSR-A, reducing in addition the probability of ionization from $\sim
30\%$ for the single BH to $\sim 8\%$ for the BH binary (Table
\ref{sample-table}, last line).
In Table \ref{sample-table} we give results
obtained for other pair values of $(m_{\rm BH},M_{\rm BH}).$ 
When modeling BH binary evolution, we find that 
the heavier BH grows in mass by capturing stars   
and the range considered bracket possible values of 
$(m_{\rm BH},M_{\rm BH})$ (Colpi, Mapelli \& Possenti 2003, in prep.).
The lightest pair can in principle simulate
ejection
of PSR-A by a binary comprising a BH and a neutron star, a possibility
that is worth mentioning. 

\section {Linking pulsar's accelerations with the BH(s) hypothesis}

The single intermediate-mass BH (of \ref{BHsingle})
can  account for
a significant portion of the unseen low luminosity matter required for
explaining the spin derivatives of PSR-B, PSR-E (and perhaps PSR-D). 
Given the cluster distance, the effects of a $500
M_\odot$ BH on the star density profile would be confined within the
inner $r_{\rm BH}/d\sim 1''$, 
thus resulting unobservable even when using  the accurate
profile of Ferraro et al. (2003), derived from HST observations.

Alternatively, one may wonder whether a suitably located perturber
of lower mass, such as a BH binary (discussed in \ref{BHbinary}) 
can contemporarily accelerate  the two pulsars.
Given
the relative projected positions of PSR-B and PSR-E, the mass of the
perturber should be close to  $100~{\rm M_\odot}.$ If the BH binary  
resides close to $C_{\rm grav}$, 
dynamical encounters of the type explored in this paper 
would displace its center-of-mass causing its wandering in the core. 
In particular, assuming  an harmonic
potential for the central region of the
cluster (and a central density of  $\sim 10^5~{\rm
M_\odot~pc^{-3}}$), the minimum recoil velocity for moving 
the perturber from the
center of the potential well to the pulsar locations is $V_{\rm
min}\sim 4$ km s$^{-1}.$ Table \ref{sample-table}
shows that a binary BH of the required  mass  has typical
$V_{\rm BH}\sim 1.5\pm1.0$ km s$^{-1}$ following a dynamical
encounter. Hence reaching the pulsar positions would be possible
only if an unusually  strong interaction occurs capable of imprinting
a larger recoil velocity.

\begin{table}
\caption{Binary-Binary Encounters}
\tabcolsep 5pt
\begin{tabular}{l|cccc}
\hline 
\hline

End-states    &  $(3,30)$ & $(10,50)$ & $(10,200)$ & $(1,500)$ \\
              & $({\rm{M_\odot,M_\odot}})$ & $({\rm{M_\odot,M_\odot}})$ &
              $({\rm{M_\odot,M_\odot}})$  & $({\rm{M_\odot,M_\odot}})$ \\
\hline
$\langle V\rangle_{\rm PSR-A}({\rm km \,s^{-1}})$ & $34.7\pm 29.0$
       & $73.0\pm 53.0$ & $101.4 \pm 61.5$  & $54.6 \pm 19.1$ \\
$\langle V\rangle_{\rm BHs} ({\rm km \,s^{-1}}) $ & 
       $1.9\pm 1.6$    & $2.1\pm 1.6$ & $0.8\pm 0.6$ & $ 0.3\pm 0.3$\\
$\rm{e_{\rm\,\,peak,\,\, PSR-A}}$ & $9.4\times 10^{-6}$ &
       $1.9\times 10^{-5}$ & $1.6\times 10^{-4}$ & $1.2\times 10^{-2}$\\
$\% \rm \,\,\,\,{ionizations}$ & $8.3\%$ & $10.7\%$ & $12.1\%$ & $27.3\%$ \\
\hline \hline
\end{tabular}
\label{sample-table}
\end{table}

\end{document}